\documentclass[twocolumn,showpacs,preprintnumbers,amsmath,xamssymb,aps,pop]{revtex4-1}

\usepackage{amssymb,amsfonts,amstext,graphics,graphicx,subfigure}
\usepackage[colorlinks]{hyperref}
\usepackage{color}

\usepackage{mathbbol}

\begin{document}

\title{Metriplectic torque for rotation control of a rigid body}
\author{Massimo Materassi}
\email{massimo.materassi@isc.cnr.it}
\affiliation{Istituto dei Sistemi Complessi, Consiglio Nazionale delle Ricerche, Italy}
\author{Philip J. Morrison}
\email{morrison@physics.utexas.edu}
\affiliation{Department of Physics and Institute for Fusion Studies, 
The University of Texas at Austin, Austin, TX, 78712, USA}
\date{\today}

\begin{abstract}

Metriplectic dynamics  couple a Poisson bracket of the Hamiltonian description with a kind of metric bracket,  for describing systems with both Hamiltonian and dissipative components. The construction builds in asymptotic convergence to a preselected equilibrium state. Phenomena such as friction, electric resistivity, thermal conductivity and collisions in kinetic theories all fit within this framework.  In this paper an application of metriplectic dynamics  is presented that is of interest for the theory of control: a suitably chosen torque, expressed through a {metriplectic extension of its ``€œnatural" Poisson algebra}, an algebra obtained by reduction of a canonical Hamiltonian system, is applied to a free rigid body.  On a practical ground, the  effect is to drive the body to align its angular velocity to rotation about a stable principal axis of inertia, while conserving its kinetic energy in the process. On theoretical grounds, this example provides a class of non-Hamiltonian torques  that  can be added to the canonical Hamiltonian description of the free rigid body and  reduce to metriplectic dissipation.  In the canonical description these torques   provide convergence to a higher dimensional attractor.  The method of construction  of such torques can be extended to other dynamical systems describing ``machines"€ with non-Hamiltonian motion  having  attractors.

\end{abstract}

\maketitle


\section{Introduction}
This work is about a new feature of  metriplectic dynamics \cite{Morrison86}, an extension of the Hamiltonian formalism with  dissipation that induces time-asymptotic convergence to equilibrium solutions. This convergence is realized in metriplectic dynamics  by adding to the antisymmetric Poisson bracket a bilinear symmetric bracket with desirable properties. The resulting  function algebra of observables, which will be seen to generate the evolution, satisfies the Leibniz derivation property with pointwise multiplication. 

In particular,  in metriplectic dynamics   the total energy of the system, namely the Hamiltonian $H$, is conserved, while another functional, referred to as the entropy $S$, grows to its maximum.  Thus, the dynamics effects the free-energy-like variational principle
\begin{equation}
\delta F= \delta (H+\zeta S)=0
\label{VP}
\end{equation}
 ($\zeta$ a Lagrange multiplier) converging  asymptotically to a desired equilibrium state.

The metriplectic  formalism has been identified for many finite and infinite-dimensional cases in which a Hamiltonian physical system is coupled to €œmicroscopic degrees of freedom, € giving rise to dissipation.  For example,  the addition of the effects of collisions in the Vlasov kinetic theory  \cite{Morrison84a} and effects of  \emph{friction} and  \emph{thermal conduction} in fluid   \cite{Morrison84b} and magnetohydrodynamic  \cite{MateTassi12} systems have been obtained.

Here we focus on a particular finite-dimensional case, already introduced in \cite{Morrison86}, in which the Hamiltonian system representing a free rigid body is perturbed with an external torque $\vec{\tau}_{\mathrm{servo}}$ suitably \emph{designed to modify the angular momentum $\vec{L}$ without changing the energy of the system}. In particular, the action of this torque makes the angular velocity $\vec{\omega}$ converge to a free rotation around one principal axis.    

The construction of \cite{Morrison86} is based on the invariance of $L^2$,  a so-called Casimir invariant of  the rigid body Poisson bracket that produces the Hamiltonian part of the system. As the torque applied must be a suitable function of the angular momentum $\vec{\tau}_{\mathrm{servo}} = \vec{\tau}_{\mathrm{servo}}(\vec{L})$, the technological solution imagined is a servo-motor, which we term a \emph{metriplectic servo-motor} (MSM).  The intriguing aspect of all this  is that, due to the energy conservation, the MSM would re-direct  without any power consumption, as $\vec{\tau}_{\mathrm{servo}} \cdot \vec{\omega} = 0$. 

The metriplectic rigid body system can be viewed in two ways:  in terms of the three-dimensional Euler equations, the reduced system  with variables $\vec{\omega}$ as in  \cite{Morrison86},  or in terms of the unreduced system, the  six-dimensional  Hamiltonian system with  canonical variables $\left(\chi,\vec{p}\right)$, where $\chi$ denotes the Euler angles and $\vec{p}$ their canonically conjugate momenta.  It is this latter system that  is presented here for the first time.

The metriplectic dynamics  of  \cite{Morrison86} in terms of  $\vec{\omega}$ relaxes to an asymptotic state of free rotation around a stable  principal axis  and thus  has a point attractor.  However, the system we propose here in terms of the  $\left(\chi,\vec{p}\right)$ relaxes  to  a  cylinder, higher dimensional attractor.  This observation leads to a systematic method for appending non-Hamiltonian terms to a canonical Hamiltonian system that give rise to systems with attractors.  A correspondence between the Lie-Poisson systems,  the standard classical form for macroscopic models of matter obtained upon reduction of canonical Hamiltonian systems (see e.g.\  \cite{Morrison98}), with the addition of metriplectic dissipation and their unreduced canonical systems with the addition of non-Hamiltonian yet reducible terms is made.

The paper is organized as follows. In Section \ref{sec:metrisys} a short review of metriplectic dynamics  is presented. In Section \ref{sec:servomotor} the application of the MSM to the free rigid body is described, first in terms of the phase space with the angular velocity $\vec{\omega}$ (or equivalently $\vec{L}$) as coordinates, then in Section \ref{sec:limitcyc} in terms of  the canonical variables $\left(\chi,\vec{p}\right)$.  In these sections it  is shown that the rotation around a principal axis of inertia corresponds to an asymptotic equilibrium point in the $\vec{\omega}$ space, while it is an attracting cylinder of periodic orbits in the canonical variables $\left(\chi,\vec{p}\right)$.  Section \ref{sec:servomotor}  includes numerical examples.  Conclusions and further development of the present study, both in physical and technological senses, are drawn in Section \ref{sec:conclusion}.


\section{A Review of Metriplectic Dynamics} 
 \label{sec:metrisys}

As noted above, metriplectic  dynamical systems are an extension of the usual Poisson bracket formulation  of  Hamiltonian systems. The aim of this extension is to obtain systems with a  relaxation processes that conserves the Hamiltonian $H$ while increasing an ``entropy" $S$ until an asymptotically stable equilibrium is reached.  Thus,  metriplectic dynamics  provides a dynamical realization of laws of thermodynamics: the  first law by conserving $H$, and the second law  by increasing $S$ until an equilibrium is reached.  

Typically, one starts with a noncanonical Hamiltonian system of Hamiltonian $H(z)$ and a degenerate Poisson bracket $\left\{ .,.\right\} $, $z$ being the dynamical variables that are coordinates of a phase space that is a Poisson manifold (see e.g.\ \cite{Morrison98} for review). The evolution of any observable $A$ is given by 
\begin{equation}
\dot{A}\left(z\right)=\left\{ A\left(z\right),H\left(z\right)\right\}.
\label{eq:01.Adot.Hamiltonian}
\end{equation}
Because  $\left\{ .,.\right\}$ is antisymmetric, if follows that $H$ is conserved, along with usual Noether invariants $I$ that satisfy $\left\{H,I\right\}=0$ for the specific $H$ of the system.   For noncanonical Hamiltonian systems with degenerate Poisson brackets,  there may be another invariant that has a null Poisson bracket with {\em any} other function of $z$; i.e., 
\begin{equation}
\left\{ S\left(z\right),A\left(z\right)\right\} =0\quad \forall\  A.
\end{equation}
An invariant $S$ of this type is called a  Casimir invariant of the Poisson bracket.  Observe, it is built into the Poisson bracket, and  is necessarily conserved by the dynamics of (\ref{eq:01.Adot.Hamiltonian}) for any  Hamiltonian one might have.   In metriplectic dynamics  Casimir invariants are candidate entropies, the choice of which determines the variational principle of \eqref{VP} and consequently the state to which the system relaxes. 

In metriplectic dynamics  the increase of entropy is assured by employing a symmetric bilinear  bracket satisfying
\begin{equation}
\left(A,B\right)=\left(B,A\right)\quad \mathrm{and}\quad  \left(A,A\right)\le0\quad \forall\  A,B.
\label{eq:02.ABsymm.def}
\end{equation}
In addition,   the invariance of  $H$ is assured by the following degeneracy condition:
\begin{equation}
\left(H,A\right)=0\quad \forall\  A.
\label{eq:03.HsymmA.zero}
\end{equation}
Just as Poisson brackets satisfy the Leibniz rule, we assume our symmetric bracket acts as a derivation in each argument. 

The metriplectic dynamics that  determines the evolution of  an observable $A$  is given in terms of a  metriplectic bracket
\begin{equation}
\left\langle \left\langle A,B\right\rangle \right\rangle =\left\{ A,B\right\} +\left(A,B\right)
\end{equation}
as 
\begin{eqnarray}
\label{eq:04.Adot.CMS}
\dot{A}\left(z\right)&=&\left\langle \left\langle A(z),F(z)\right\rangle \right\rangle
\\
&=&\left\{ A\left(z\right),H\left(z\right)\right\} +\zeta\left(A\left(z\right),S\left(z\right)\right).
\nonumber
\end{eqnarray}
where the second equality follows from the properties above and the definition $F=H +\zeta S$.
We refer the reader to Section 4 of \cite{BMR13} for a more rigorous mathematical definition of metriplectic dynamics  along with  background references.  

To reiterate, it is clear from the above properties that the dynamical system $\dot{z}= \langle  \langle z,F(z)\rangle\rangle$ will satisfy
\begin{equation}
\dot{H}\left(z\right)=0 \quad \mathrm{and}\quad \dot{S}\left(z\right)\ge0;
\label{eq:1st.2nd.principles}
\end{equation}
with the inequality of \eqref{eq:1st.2nd.principles} following from $\dot{S}=\zeta\left(S,S\right)$ in light of \eqref{eq:02.ABsymm.def},  provided $\zeta$ is chosen to be negative.  Thus  the Casimir invariant $S$ plays the role of a \emph{ Lyapunov functional} and the asymptotic equilibrium of $\dot{z}= \langle  \langle z,F(z)\rangle\rangle$  is the extremum of the free energy $F$.  The terminology free energy for $F$ is natural, given the equilibrium thermodynamic analogy.

The review of this section was written so as  not to distinguished between the metriplectic dynamics  of finite-dimensional and infinite-dimensional systems (field theories).  In the following section, an  explicit finite-dimensional example, a set of ordinary differential  equations  for the free rigid body,   will be  presented.  This example illustrates  the main point of the paper:  how to amend a canonical Hamiltonian system in such a way that  that it possesses relaxation  to a limit cycle.


\section{The metriplectic servo-motor for rigid body}
\label{sec:servomotor}

Now consider our example, a rigid body of inertia tensor $\sigma$ with the three eigenvalues, principal moments of inertia,  $\left\{  I_1, I_2, I_3 \right\} $. As noted in the Introduction, the rigid body may be described via a phase space with canonical coordinates $\left( \chi, \vec{p} \right)$,  the set of the three Euler angles $\chi$ and their conjugate atmomenta $\vec{p}$. 
 In the absence of external torques, its Hamiltonian is merely its kinetic energy, 
\begin{equation}
H\left(\chi,\vec{p}\right)=\frac{1}{2}\vec{L}^{\mathrm{T}}\left(\chi,\vec{p}\right)\cdot\sigma^{-1}\cdot\vec{L}\left(\chi,\vec{p}\right),\label{eq:H.rigidbody.Chi.P}
\end{equation}
where recall  $\vec{L}$ is the angular momentum of the rigid body. The three components of $\vec{L}$ form a closed noncanonical Poisson algebra, given by the Poisson tensor
\begin{equation}
\left\{ L_{i},L_{j}\right\} =-    \epsilon_{ij}^{\phantom{ij}k} L_{k}\,,
\label{eq:Li.Poisson.Lj}
\end{equation}
where $ \epsilon_{ij}^{\phantom{ij}k}$ is the purely antisymmetric  Levi-Civita tensor and repeated summation over $k$ is implied.    Equation \eqref{eq:Li.Poisson.Lj} defines the Lie-Poisson bracket defined on functions of $\vec{L}$ associated with the Lie algebra of rotations $\mathfrak{so}(3)$.  Because both the Poisson bracket and the Hamiltonian can be written in terms of $\vec{L}$, the six dimensional system in terms of $(\chi,\vec{p})$ with the canonical Poisson tensor $\left\{ \chi_{i},p_{j}\right\} = \delta_{ij}$, reduces to the three-dimensional one  in terms of $\vec{L}$ alone, with the Hamiltonian
\begin{equation}
H\left(\vec{L}\right)=\frac{1}{2}\vec{L}^{\mathrm{T}} \cdot\sigma^{-1}\cdot\vec{L},
\label{eq:H.rigidbody.L}
\end{equation}
and the Poisson tensor given by \eqref{eq:Li.Poisson.Lj}. (See e.g.\ \cite{Morrison98} for review.)
In the reduced system the evolution of any observable $A(\vec{L})$ have the  noncanonical Hamiltonian dynamics,  
\begin{eqnarray}
\dot{A}\left(\vec{L}\right)&=&\left\{ A\left(\vec{L}\right),H\left(\vec{L}\right)\right\} 
\nonumber\\
&=& \frac{\partial A}{\partial L_i} \{L_i,L_j\} \frac{\partial H}{\partial L_j}\,.
\label{eq:dotA.Hamiltonian.L}
\end{eqnarray}
Upon choosing the observable $A$ to be  $\vec{L}$.   \eqref{eq:dotA.Hamiltonian.L}  yields Euler's  equations for the free rigid body, 

Due to the form of the Poisson bracket defined by  (\ref{eq:Li.Poisson.Lj}), any function $C ( L^2  )$ of the square modulus of $\vec{L}$ is a Casimir invariant, i.e.
\begin{eqnarray}
\left\{ C\left(L^{2}\right),A\left(\vec{L}\right)\right\} &=&\frac{\partial C}{\partial L^{2}}\left\{ L^{2},A\left(\vec{L}\right)\right\}
\nonumber\\
&=& 0\quad \forall\  A.
\end{eqnarray}
Thus, under the dynamics of  \eqref{eq:dotA.Hamiltonian.L}  the Casimir  $C( L^2 )$ is  conserved  and even if one altered $H( \vec{L})$ by the addition of  any function of $L^2$, the evolution would  remain unchanged.   Thus, $C( L^2 )$ is available to be the  entropy of our metriplectic system.

In order to construct a metriplectic system out of the Hamiltonian system (\ref{eq:Li.Poisson.Lj}) and (\ref{eq:H.rigidbody.L}), one that converges to a free rotation around one of its principal axes as an asymptoticly stable equilibrium while conserving $H$, a bracket of the general form
\begin{equation}
\left(A,B\right)=\Gamma^{ij}\frac{\partial A}{\partial L^{i}}\frac{\partial B}{\partial L^{j}}
 \label{eq:AsymmB.Gamma}
\end{equation}
with a certain symmetric semi-definite tensor $\Gamma ( \vec{L} )$ must be  tailored to our needs.  In particular, for energy conservation we require $\Gamma$ have the degeneracy condition that ensures $\Gamma \cdot {\partial H}/{\partial \vec{L}} = 0$ for the $H$ of \eqref{eq:H.rigidbody.L}. The simplest possible choice for the tensor $\Gamma$ was chosen in \cite{Morrison86},  
\begin{equation}
\Gamma\left(\vec{L}\right)=\frac{k}{\zeta}\left(\omega^{2}\boldsymbol{1} 
-\vec{\omega}\otimes\vec{\omega}\right)
\label{eq:Gamma.Rigidbody.01}
\end{equation}
where 
\begin{equation}
 \vec{\omega}=\frac{\partial H}{\partial\vec{L}}=\sigma^{-1}\cdot\vec{L}
 \label{omegaL}
\end{equation}
is    the angular velocity of the rigid body.  In (\ref{eq:Gamma.Rigidbody.01}) $\Gamma$ is proportional to the projector perpendicular to ${\partial H}/{\partial \vec{L}}$; consequently,  this choice fulfills the prescription of (\ref{eq:03.HsymmA.zero}).

With the above Poisson and symmetric bracket choices,  the metriplectic dynamics of  an observable $A(\vec{L})$ is given by  
\begin{eqnarray}
\dot{A}\left(\vec{L}\right)&=&\left\langle \left\langle A\left(\vec{L}\right),F\left(\vec{L}\right)\right\rangle \right\rangle
 \label{eq:dotA.rigidbody.CMS}\\
&=&\left\{ A\left(\vec{L}\right),H\left(\vec{L}\right)\right\}
  +\zeta\left(A\left(\vec{L}\right),C\left(L^{2}\right)\right),
\nonumber
\end{eqnarray}
where 
\begin{equation}
F\left(\vec{L}\right)=H\left(\vec{L}\right)+\zeta C\left(L^{2}\right)
\label{eq:free.energy.F.rigidbody}
\end{equation}
is the chosen free energy.

The metriplectic dynamics  of (\ref{eq:dotA.rigidbody.CMS}) with \eqref{eq:free.energy.F.rigidbody} conserves $H$ and varies $C$ monotonically, until the free energy of  \eqref{eq:free.energy.F.rigidbody} reaches its extremum, ${\partial F}/{\partial \vec{L}} = 0$. This is a relaxation process to a configuration where the rigid body rotates about an inertial axis, because the extremum of  $F( \vec{L} )$ is realized for $\vec{\omega} = -2 \zeta C'( L^2 ) \vec{L}$, with  $C'= {\partial C}/{\partial L^2}$, i.e. \emph{when the angular velocity of the rigid body and its angular momentum are aligned}, viz. 
\begin{equation}
\vec{\omega}_{\mathrm{eq}}=I_{\mathrm{eq}}\vec{L}_{\mathrm{eq}}
\quad\mathrm{where}\quad I_{\mathrm{eq}}=-\zeta C'\left(L_{\mathrm{eq}}^{2}\right)\,.
 \label{eq:rigidbody.CMS.equilibrium}
\end{equation}
Clearly, as $I_\mathrm{eq}$ is a positive quantity, the factors in $\zeta C'\left(L^2\right)$ must render it positive.   Thus,  $C$ can serve as a Lyapunov function for this dynamics.

Because the  metriplectic dynamics  of (\ref{eq:dotA.rigidbody.CMS}) changes  $\vec{L}$, implicit in the dynamics is  the  equivalent of an external torque. In particular, the torque encoded in (\ref{eq:dotA.Hamiltonian.L}) is the following:
\begin{equation}
\vec{\tau}_{\mathrm{servo}}\left(\vec{L}\right)=2kC'\left(L^{2}\right)\left[\omega^{2}\vec{L}-\left(\vec{\omega}\cdot\vec{L}\right)\vec{\omega}\right].
\label{eq:tau.servo.L}
\end{equation}
This torque  must be applied from outside, but it  depends on the instantaneous state of the rigid body through $\vec{L}$    or $\vec{\omega}$ (which are simply related by \eqref{omegaL}).   In order to technologically realize  this system one would require  a servo-mechanism that constantly senses what the rigid body is precisely doing. Such a mechanism will be referred to as a  \emph{metriplectic servo-motor} (MSM).  Remarkably, because  $\vec{\tau}_\mathrm{servo} \cdot\vec{\omega} = 0$, the mechanical power of the servo-motor vanishes, so that such a MSM could drive to the alignment of (\ref{eq:rigidbody.CMS.equilibrium}) for  rigid body of any size with no power consumption, as far as the mechanical labor is concerned. Of course, as  a $\vec{\tau}_\mathrm{servo}$ depends on the condition of the rigid body, it must take some energy to measure, save,  and react to this. Still, provided friction is minimized, so that the present metriplectic dynamics   holds, the only energy cost of an engine producing the $\vec{\tau}_\mathrm{servo}$ of  (\ref{eq:tau.servo.L}) will be that of the actuation electronics measuring the angular momentum vector, calculating  the suitable servo-torque and applying it by redirecting energy.

In the phase space of the vectors $\vec{L}$, the reduced phase space, the equilibrium to which  (\ref{eq:dotA.rigidbody.CMS}) tends is a point-like attractor for  $\vec{L}$,   determined  by  
\begin{eqnarray}
{\displaystyle \frac{d\vec{L}}{dt}}&=&\vec{L}\times\left(\sigma^{-1}\cdot\vec{L}\right)
\nonumber\\
&& +2kC'\left(L^{2}\right)\left[\left(\vec{L}^{\mathrm{T}}\cdot\sigma^{-2}\cdot\vec{L}\right)\vec{L}\right.
\nonumber\\
&&
\left.-\left(\vec{L}^{\mathrm{T}}\cdot\sigma^{-1}\cdot\vec{L}\right)\sigma^{-1}\cdot\vec{L}\right].
\label{MPdotL}
\end{eqnarray}
A posteriori one can see that  if $\vec{L}$ is parallel to an  eigendirection of $\sigma$,  then $d\vec{L}/{dt}=0$. It is also possible to show that, if the eigenvalues of $\sigma$ are ordered as $I_1 > I_2 > I_3$, only the states $\vec{L}_{\left(1\right)}=\left(L,0,0\right)^{\mathrm{T}}$ and $\vec{L}_{\left(3\right)}=\left(0,0,L\right)^{\mathrm{T}}$ are  {stable equilibria}, while $\vec{L}_{\left(2\right)}=\left(0,L,0\right)^{\mathrm{T}}$ is  {unstable}. This is consistent with the stability of the free rigid body. 

An important characteristic of the metriplectic formalism is that the dynamics may be designed  to make the system relax to any  stable equilibrium.  In the present case, the Casimir $C$ can  be chosen  to make the system relax to either 
$\vec{L}_{(1)}$, or to $\vec{L}_{(3)}$. This will be illustrated in  numerical examples.

\begin{figure}[h]
\begin{center}
\includegraphics[scale=0.22]{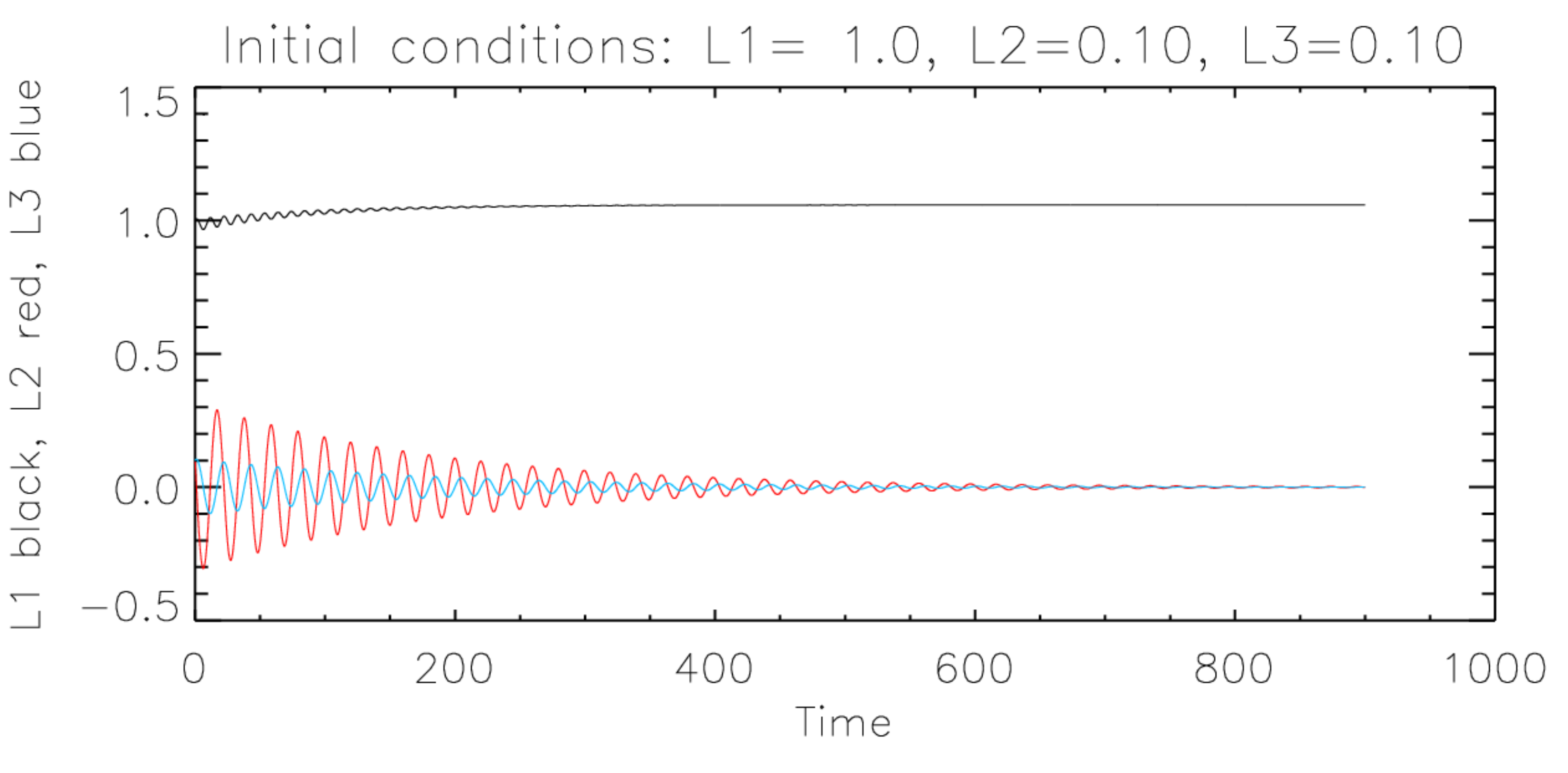}
\end{center}
\caption{Time evolution of the three components of $\vec{L}$ with the MSM with $2kC' (L^2)=0.1$ and with initial conditions $\vec{L}\left(0\right)=(1,0.1,0.1)^{\mathrm{T}}$. The system is seen to relax to $\vec{L}_{(1)}$ as designed.} \label{figure01}
\end{figure}

\begin{figure}[h]
\begin{center}
\includegraphics[scale=0.2]{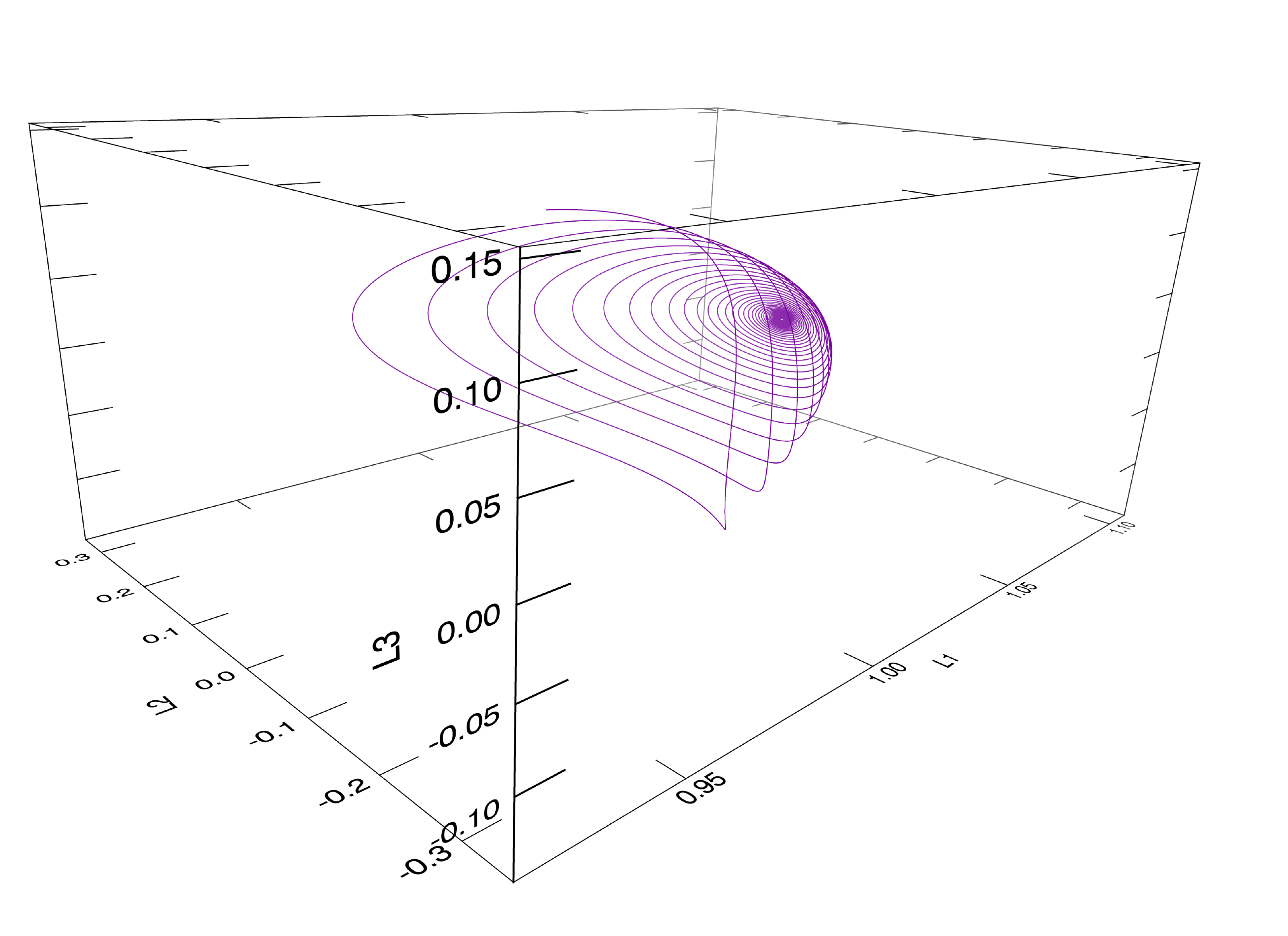}
\end{center}
\caption{Phase portrait of the MSM, with initial conditions $\vec{L}\left(0\right)=\left(1,0.1,0.1\right)^{\mathrm{T}}$, as in Figure \ref{figure01}.} \label{figure02}
\end{figure}

\begin{figure}[h]
\begin{center}
\includegraphics[scale=0.22]{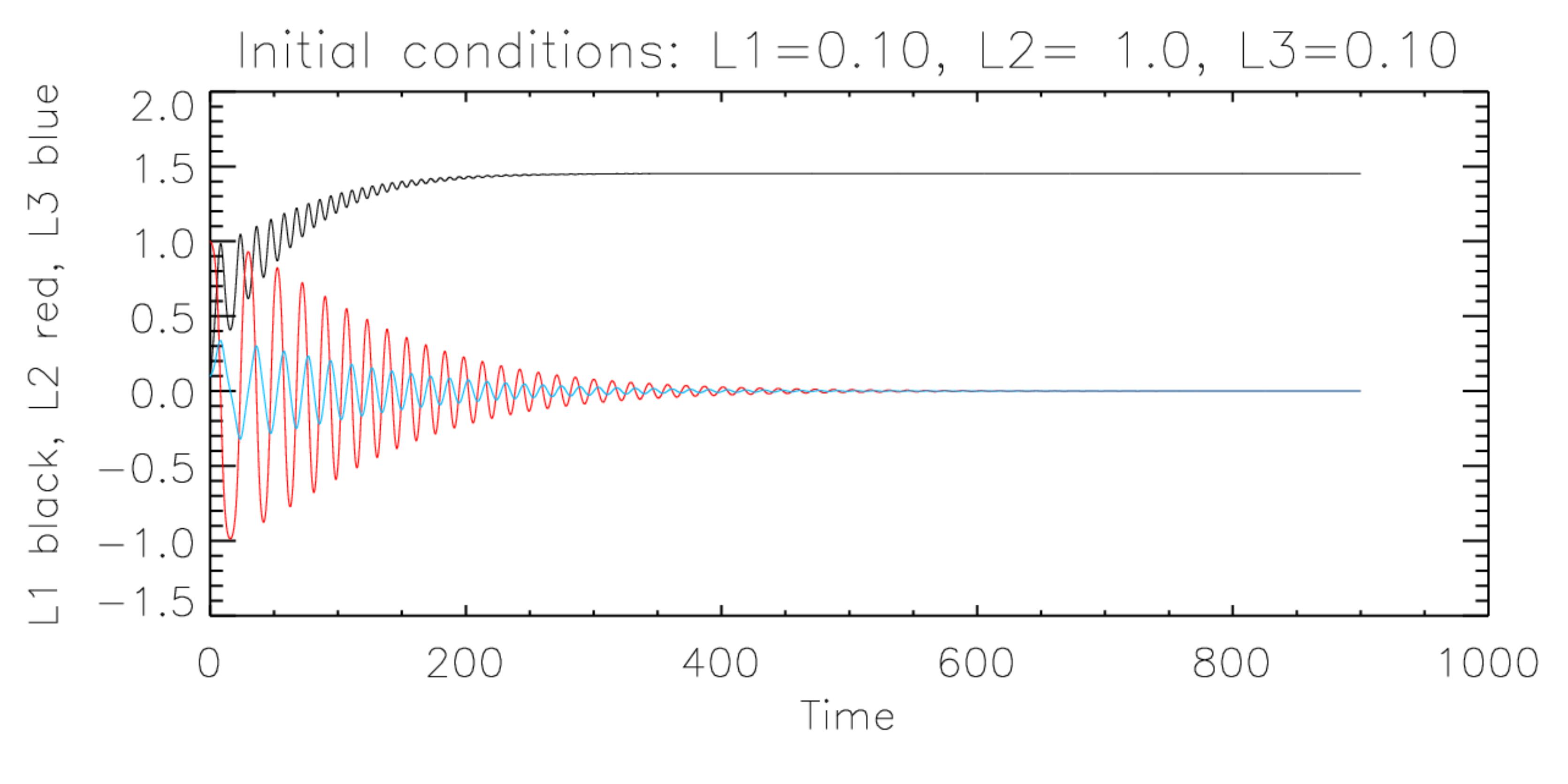}
\end{center}
\caption{Time evolution of the three components of $\vec{L}$ with the MSM with $2kC' (L^2 )=0.1$ and with initial conditions $\vec{L}\left(0\right)=\left(0.1,1,0.1\right)^{\mathrm{T}}$. The system is seen to relax to $\vec{L}_{(1)}$ as designed.} \label{figure03}
\end{figure}

\begin{figure}[h]
\begin{center}
\includegraphics[scale=0.2]{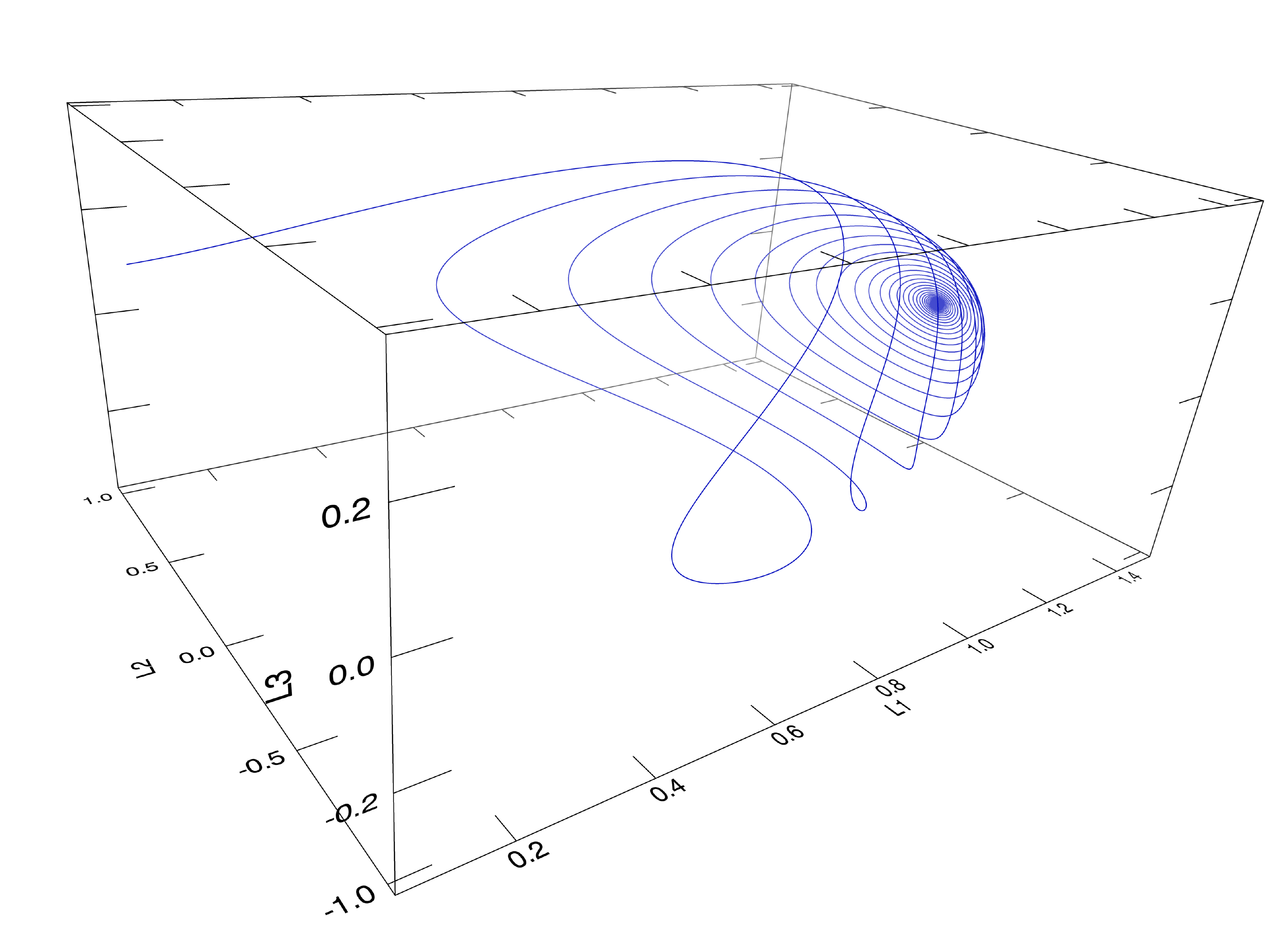}
\end{center}
\caption{Phase portrait of the MSM system, with initial conditions $\vec{L}\left(0\right)=\left(0.1,1,0.1\right)^{\mathrm{T}}$, as in Figure \ref{figure03}.} \label{figure04}
\end{figure}

\begin{figure}[h]
\begin{center}
\includegraphics[scale=0.22]{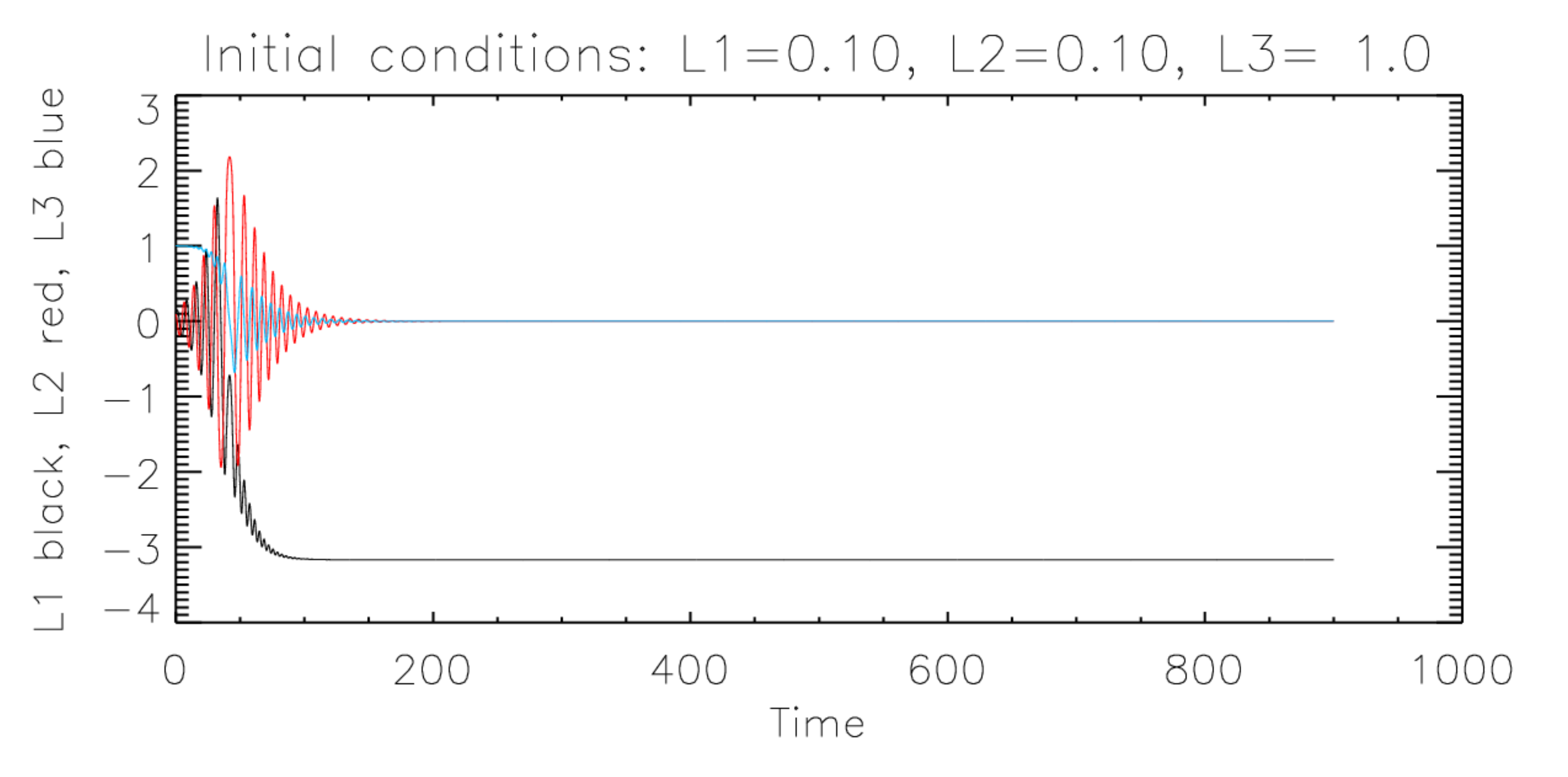}
\end{center}
\caption{Time evolution of the three components of $\vec{L}$ in the MSM system with $2kC'\left(L^2\right)=0.1$ and with initial conditions $\vec{L}\left(0\right)=\left(0.1,0.1,1\right)^{\mathrm{T}}$. The system is seen to relax to $\vec{L}_{(1)}$ as designed.} \label{figure05}
\end{figure}

\begin{figure}[h]
\begin{center}
\includegraphics[scale=0.2]{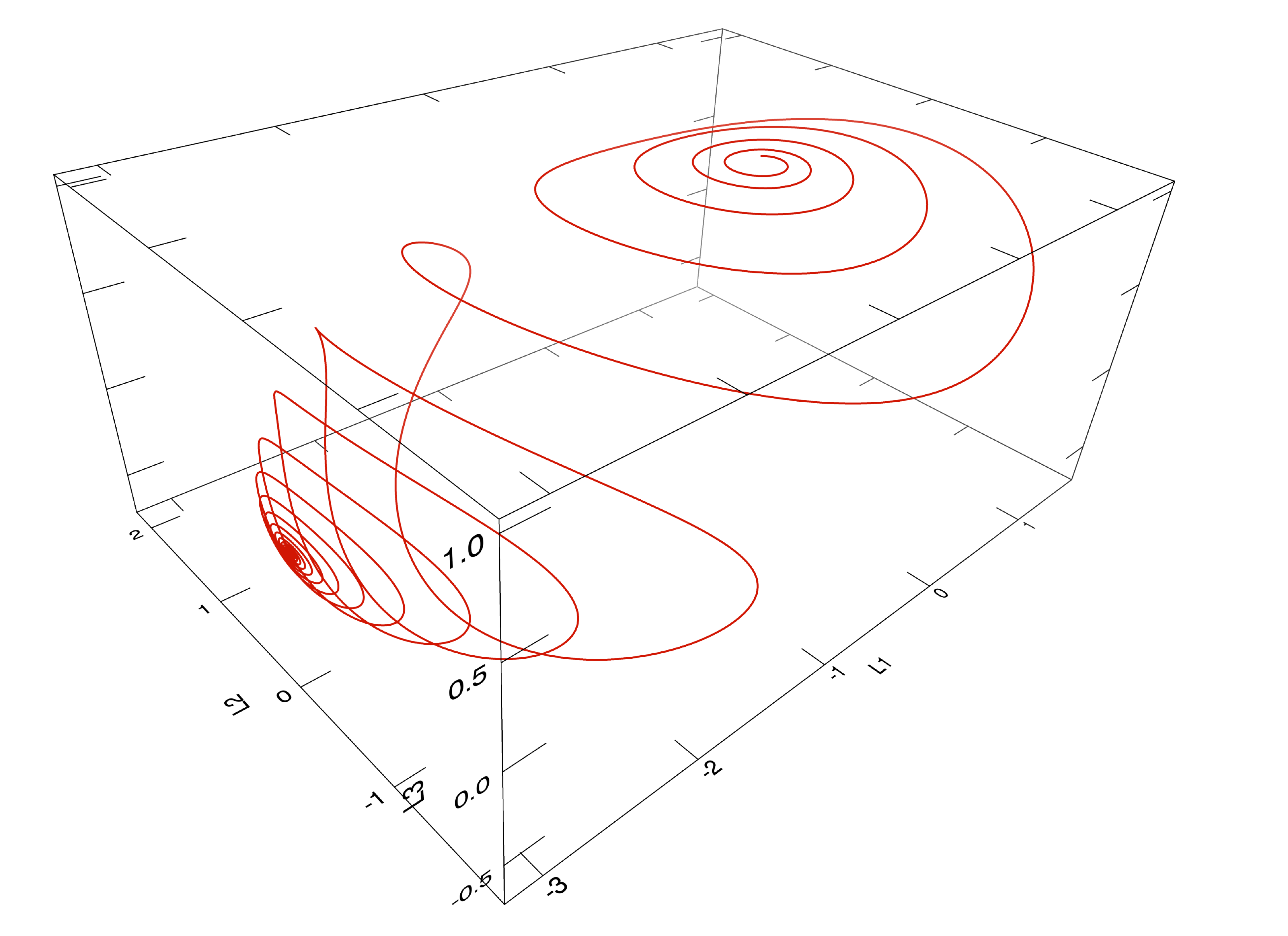}
\end{center}
\caption{Phase portrait of the MSM system, with initial conditions $\vec{L}\left(0\right)=\left(0.1,0.1,1\right)^{\mathrm{T}}$, as in Figure \ref{figure05}.} \label{figure06}
\end{figure}

Figures \ref{figure01}--\ref{figure06} depict the evolution of  the components of $\vec{L}\left(t\right)$ and the phase portrait in the $L$-space, for  the metriplectic dynamics with the MSM.  In this example, the  Casimir  $C\left(L^2\right)$ is chosen so that $2kC'\left(L^2\right)=0.1$, for different initial conditions relative to the  presumed equilibrium points.  The rigid body of moments  of inertia are set equal to $I_1 = 10$, $I_2 = 5$ and $I_3 = 1$ in arbitrary units.

\begin{figure}[h]
\begin{center}
\includegraphics[scale=0.22]{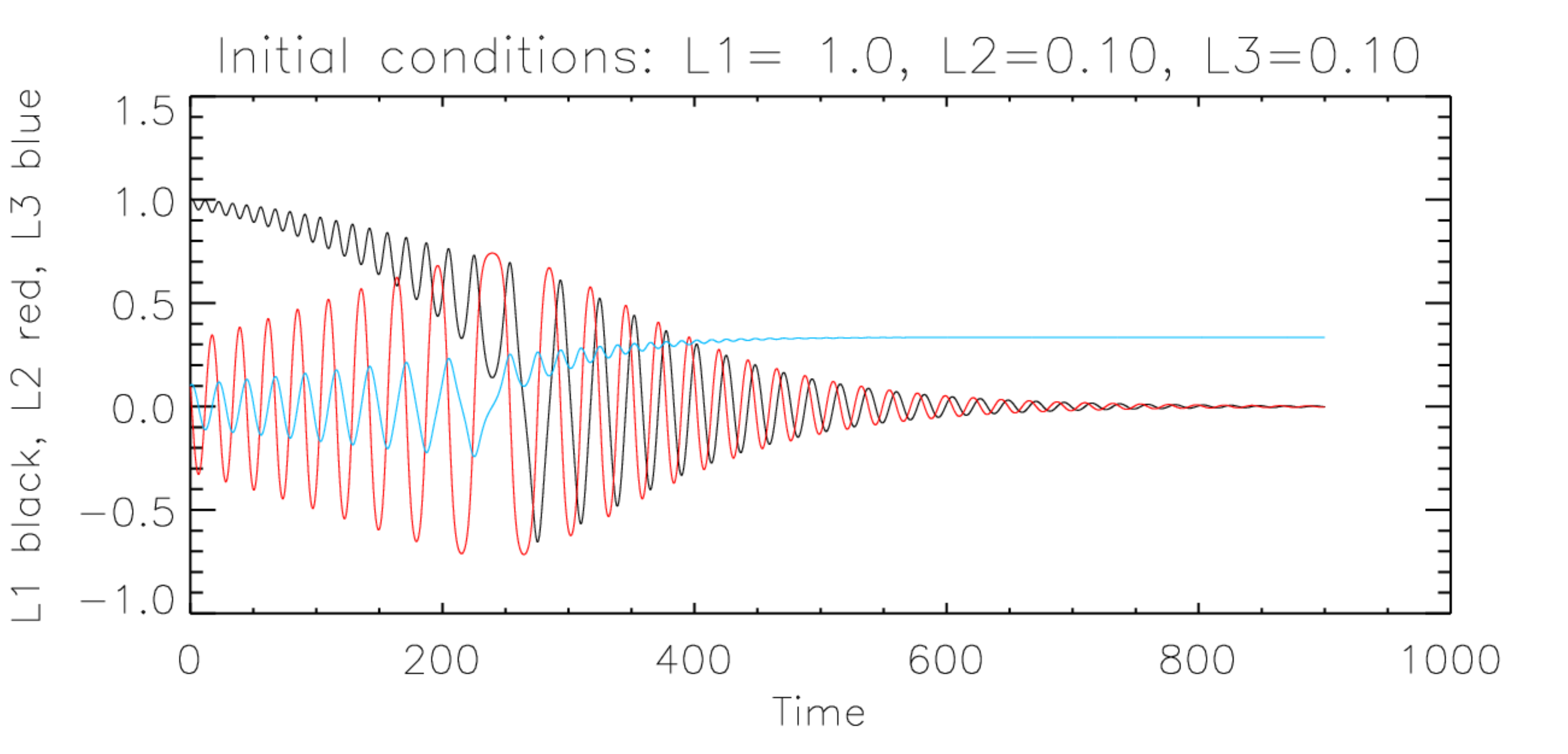}
\end{center}
\caption{Time evolution of the three components of $\vec{L}$ in the MSM system with $2kC' (L^2 )=-0.1$ and with initial conditions $\vec{L}\left(0\right)=\left(1,0.1,0.1\right)^{\mathrm{T}}$. The system is seen to relax to $\vec{L}_{(3)}$ as designed.} \label{figure07}
\end{figure}

\begin{figure}[h]
\begin{center}
\includegraphics[scale=0.2]{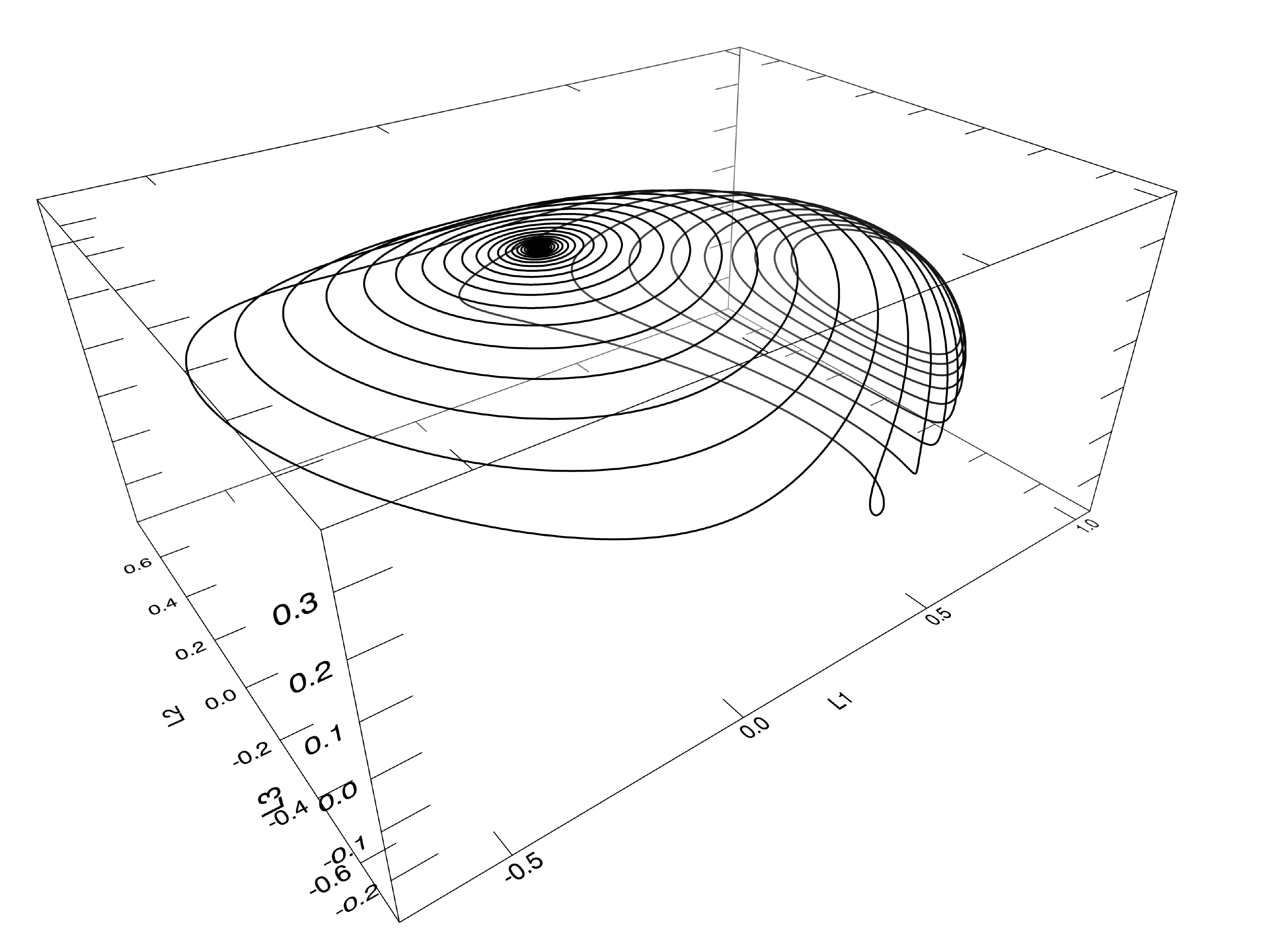}
\end{center}
\caption{Phase portrait of the MSM system, with initial conditions $\vec{L}\left(0\right)=\left(1,0.1,0.1\right)^{\mathrm{T}}$, as in Figure \ref{figure07}.} \label{figure08}
\end{figure}

\begin{figure}[h]
\begin{center}
\includegraphics[scale=0.22]{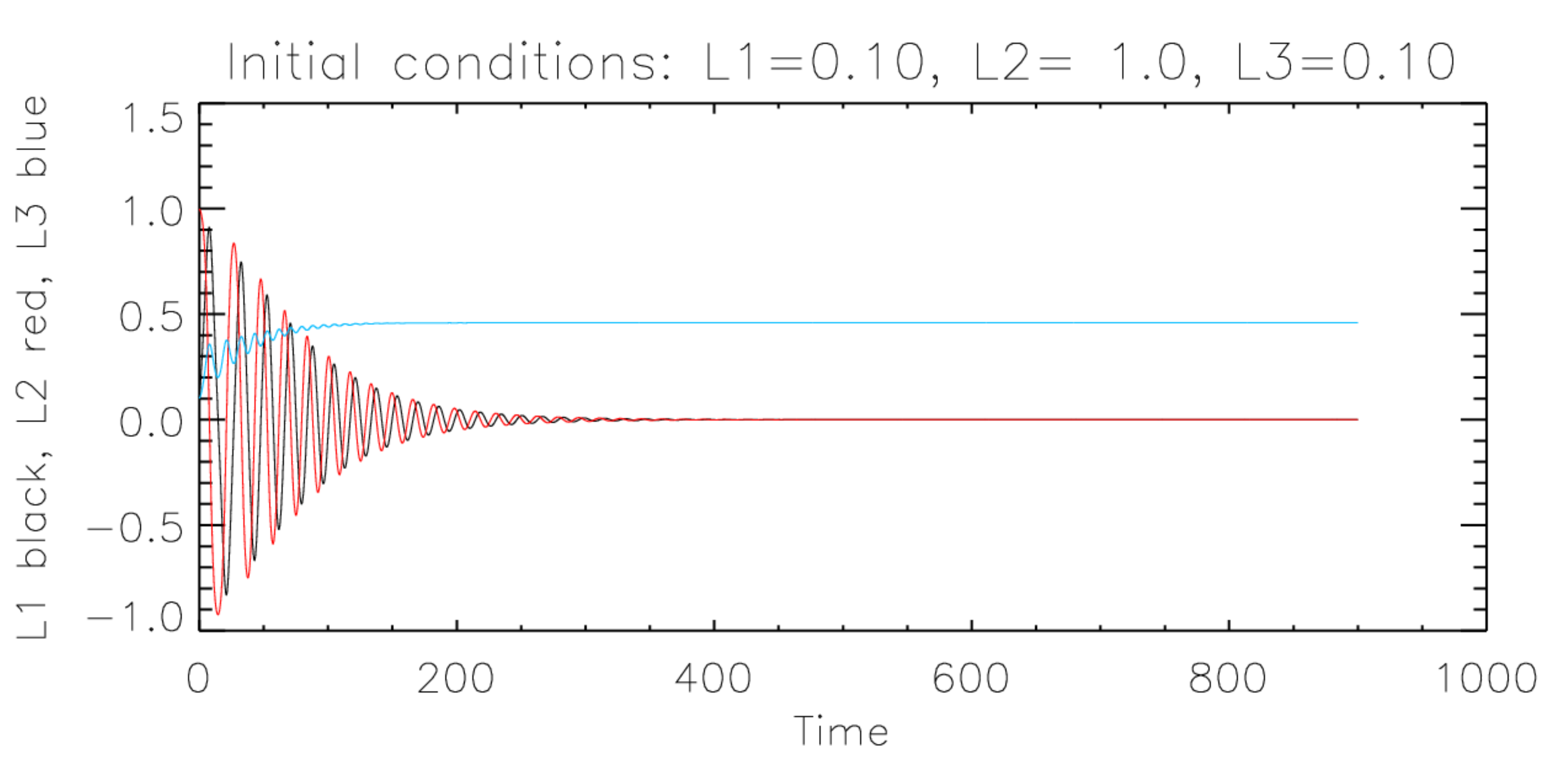}
\end{center}
\caption{Time evolution of the three components of $\vec{L}$ in the MSM system with $2kC'\left(L^2\right)=-0.1$ and with initial conditions $\vec{L}\left(0\right)=\left(0.1,1,0.1\right)^{\mathrm{T}}$. The system is seen to relax to $\vec{L}_{(3)}$ as designed.} \label{figure09}
\end{figure}

\begin{figure}[h]
\begin{center}
\includegraphics[scale=0.2]{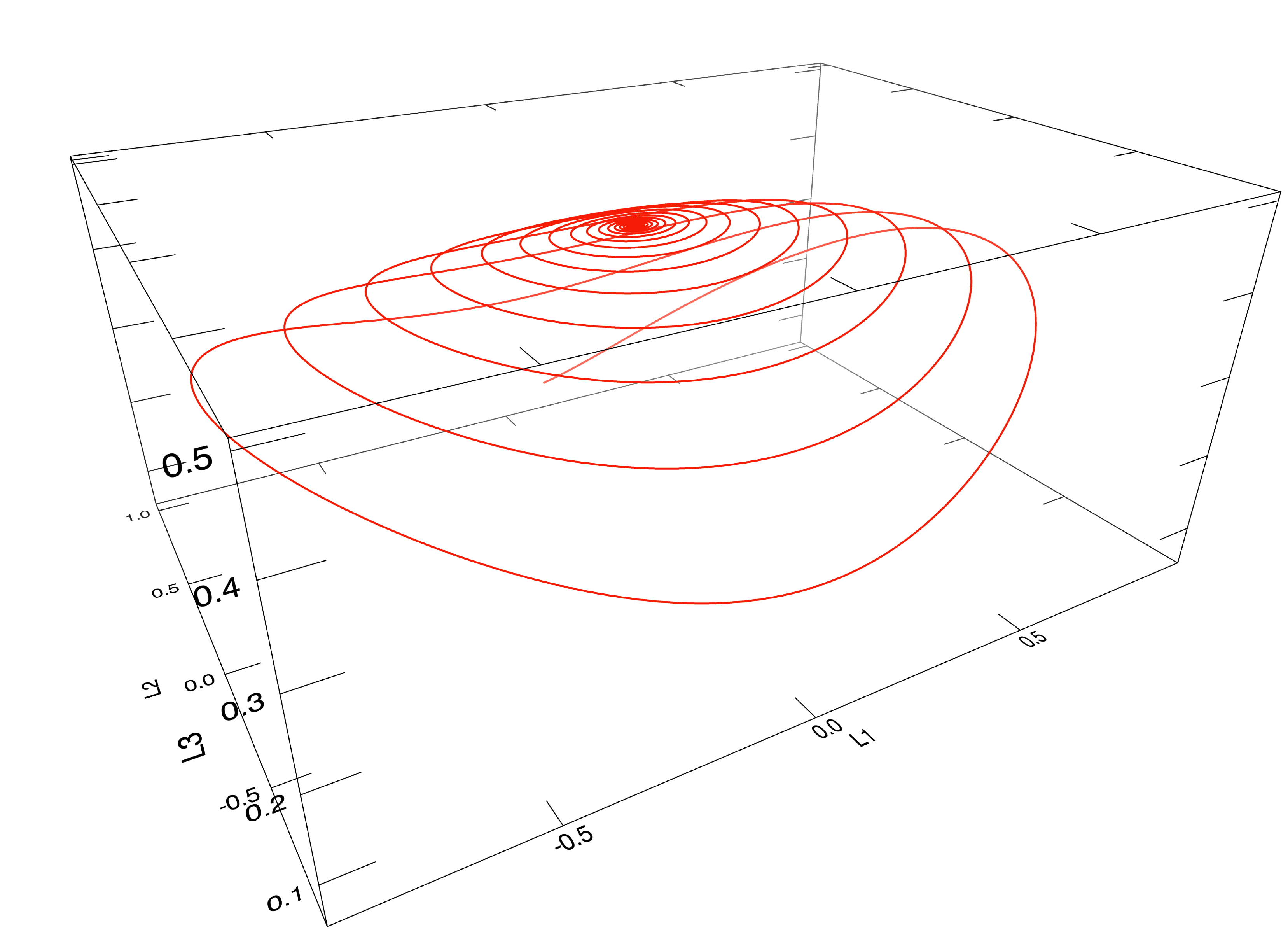}
\end{center}
\caption{Phase portrait of the MSM system, with initial conditions $\vec{L}\left(0\right)=\left(0.1,1,0.1\right)^{\mathrm{T}}$, as in Figure \ref{figure09}.} \label{figure10}
\end{figure}

\begin{figure}[h]
\begin{center}
\includegraphics[scale=0.22]{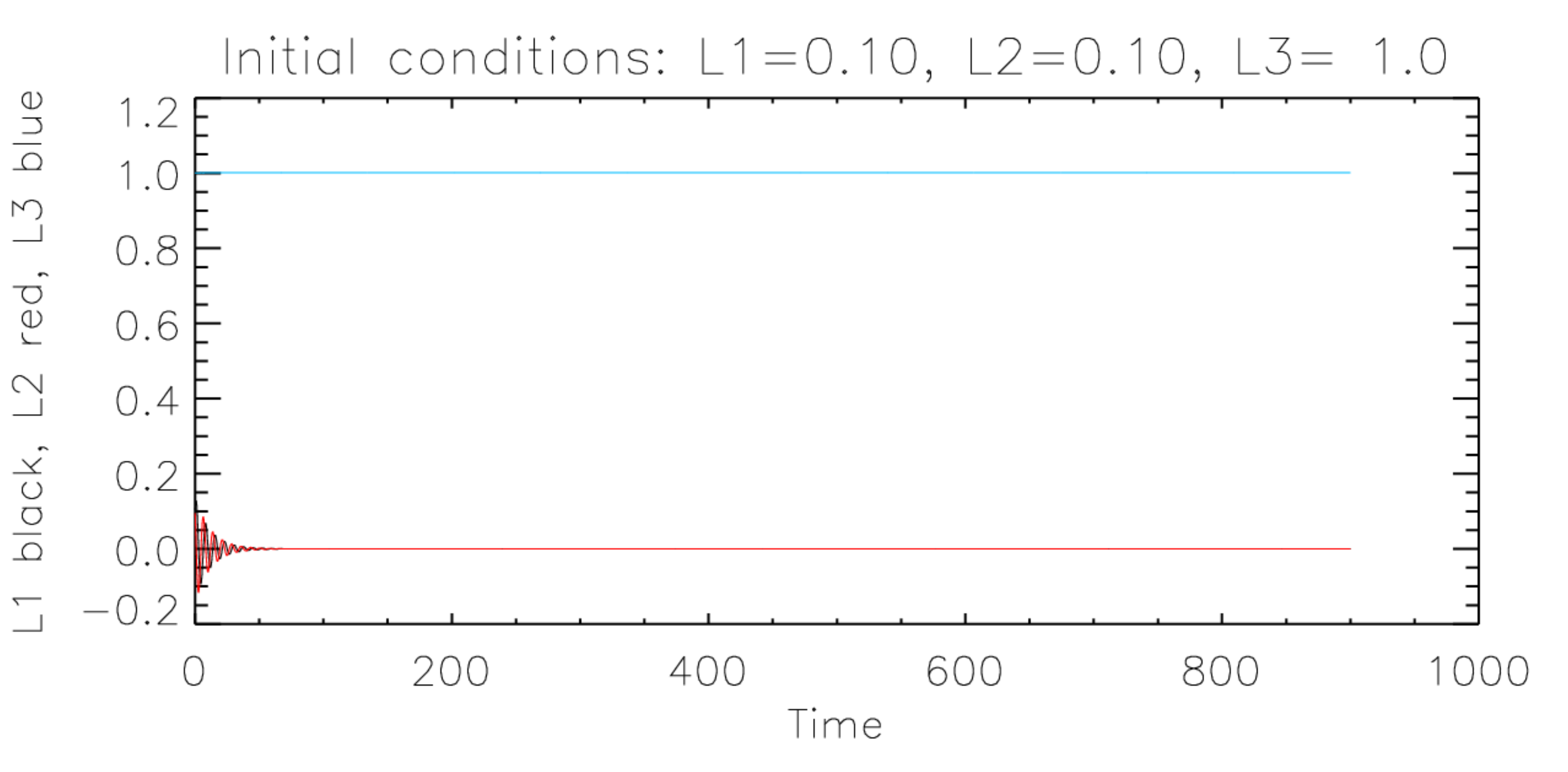}
\end{center}
\caption{Time evolution of the three components of $\vec{L}$ in the MSM system with $2kC'\left(L^2\right)=-0.1$ and with initial conditions $\vec{L}\left(0\right)=\left(0.1,0.1,1\right)^{\mathrm{T}}$. The system is seen to relax to $\vec{L}_{(3)}$ as designed.} \label{figure11}
\end{figure}

\begin{figure}[h]
\begin{center}
\includegraphics[scale=0.2]{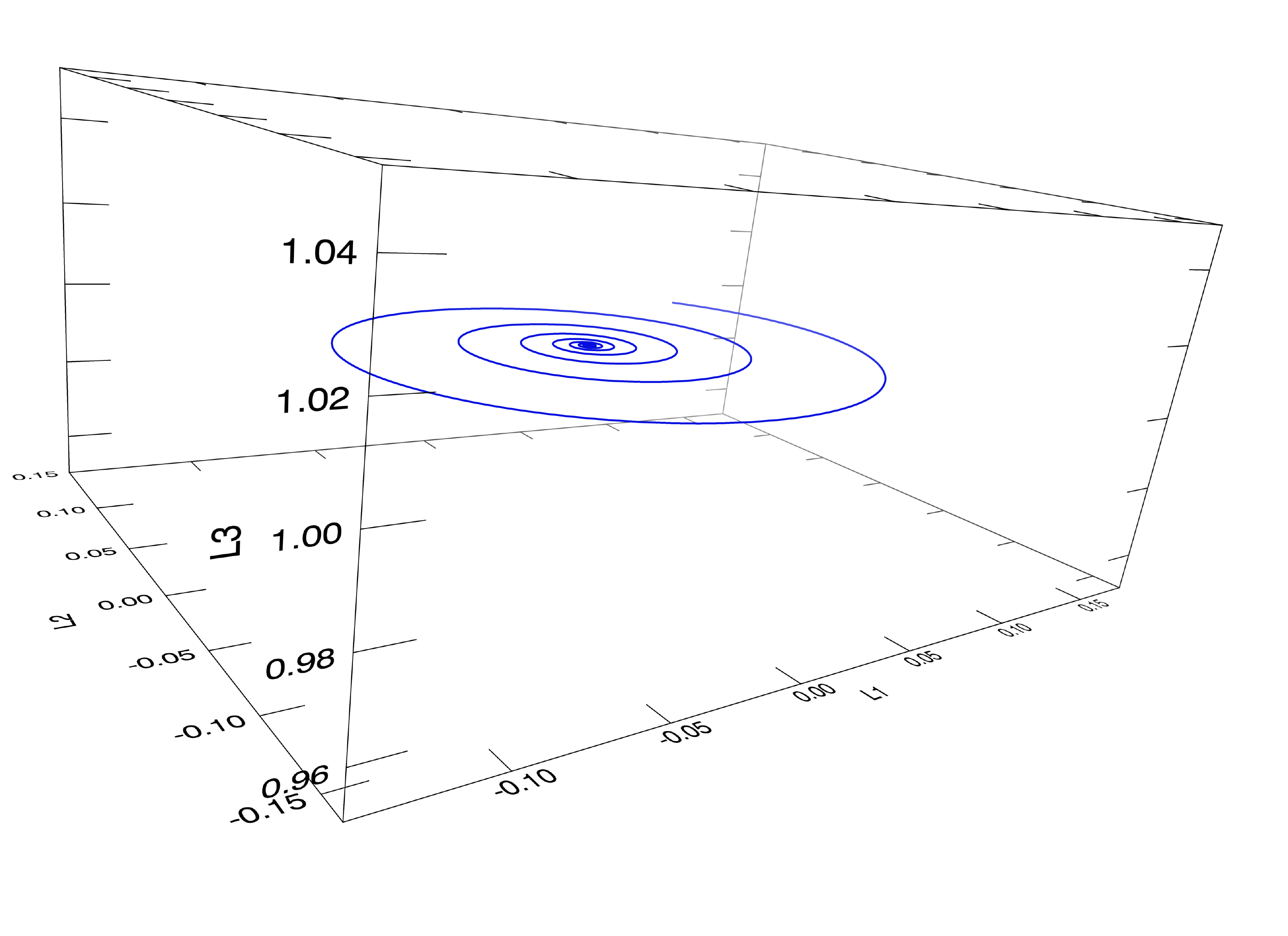}
\end{center}
\caption{Phase portrait of the MSM system, with initial conditions $\vec{L}\left(0\right)=\left(0.1,0.1,1\right)^{\mathrm{T}}$, as in Figure \ref{figure11}.} \label{figure12}
\end{figure}

Another example of MSM relaxation is shown in Figures \ref{figure07}--\ref{figure12}.  Here,   the conditions on $C\left(L^2\right)$ is changed to  $2kC'\left(L^2\right)=-0.1$, while all the remaining parameters, and initial conditions are kept the same.  The figures show that, although for  $2kC'\left(L^2\right)=0.1$ the system relaxes to $\vec{L}_{\left(1\right)}=\left(L,0,0\right)^{\mathrm{T}}$, setting  $2kC'\left(L^2\right)=-0.1$ the state $\vec{L}_{\left(3\right)}=\left(0,0,L\right)^{\mathrm{T}}$ turns out to be the global stable attractor.

\section{Attractor --  Metriplectic Reduction}
\label{sec:limitcyc}

For metriplectic systems such as the example of Section \ref{sec:servomotor}, points of asymptotic equilibria  are 0-dimensional (point) attractors,  while most interesting dissipative systems (e.g.\  electric circuits, ecological systems, etc.) have limit cycles and higher dimensional (possibly strange) attractors.  We will see, however,  that there is a correspondence between a point attractor  of a metriplectic system and a higher dimensional attractor in a larger system.

Because metriplectic systems contain noncanonical Poisson brackets with Casimirs, obtained by reduction, it is natural to investigate the dynamics in the unreduced system.   Recognizing that  the point attractors in a metriplectic systems are in fact Hamiltonian equilibrium points, for which the dissipative terms vanish, such points must correspond to solutions in the unreduced canonical Hamiltonian system.  For our  metriplectic system of Section \ref{sec:servomotor}, we will show the point attractors correspond to attracting cylinders.  

Perhaps, of even greater interest is to explore the special form of dissipative terms  that can be added to canonical Hamiltonian systems and be reduced to metriplectic systems.    Just as canonical Hamiltonian reduce to smaller noncanonical Hamiltonian systems,  we pursue metriplectic reduction, i.e.\  systems that reduce to metriplectic systems.   We will provide an example of  metriplectic reduction  by  obtaining equations in terms of $(\chi,\vec{p})$ that reduce to the metriplectic dynamics  of Section \ref{sec:servomotor}.   The equations obtained will possses the attracting cylinder.

\begin{figure}[h]
\begin{center}
\includegraphics[scale=0.2]{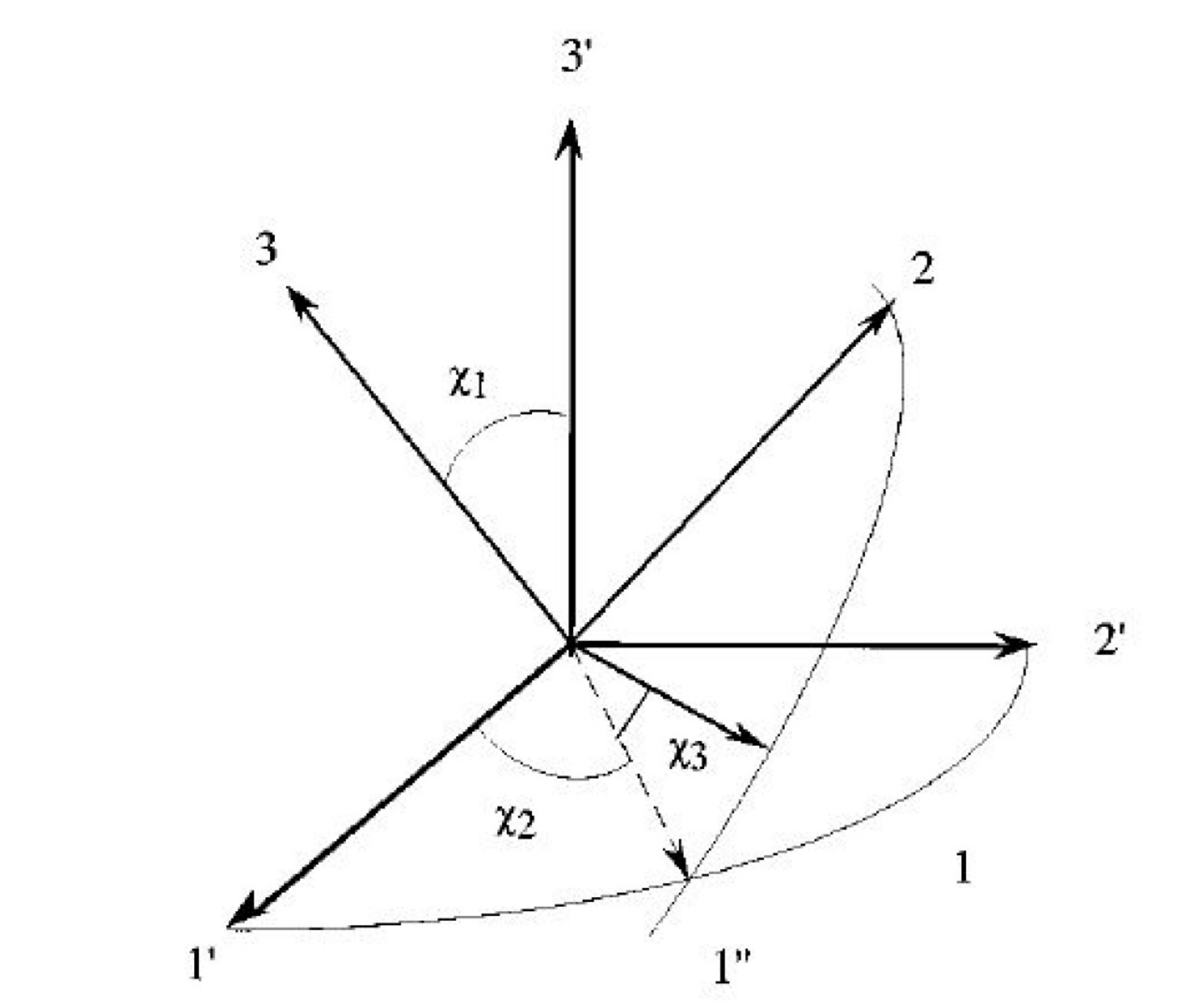}
\end{center}
\caption{The Euler angles, used as Lagrangian variables for the rigid body.}
 \label{figure13}
\end{figure}

The Hamiltonian ordinary differential equations governing the motion of the free rigid body in  the variables $\left(\chi,\vec{p}\right)$ can be found in many textbooks.  Here we use the notation of  \cite{Morrison98}, to which we will refer.  The canonical Hamiltonian equations are given by 
\begin{equation}
\begin{cases}
\quad  \vec{\dot{\chi}}&=\{\chi,H\}=\mathcal{A}^{\mathrm{T}}\left(\chi\right)\cdot\sigma\cdot \mathcal{A}\left(\chi\right)\cdot\vec{p} \,,
\\
\quad  \dot{\vec{p}}&=\{\vec{p},H\}
 \\
\quad &=-\vec{p}^{\mathrm{T}}\cdot \mathcal{A}^{\mathrm{T}}\left(\chi\right)\cdot\sigma\cdot{\displaystyle \frac{\partial \mathcal{A}\left(\chi\right)}{\partial\chi}}\cdot\vec{p}\,,
\end{cases}
 \label{eq:eq.016.ODE.chi.p.Hamiltonian}
\end{equation}
where the various contractions are  evident upon making use of the canonical Poisson bracket and Hamiltonian of \eqref{eq:H.rigidbody.Chi.P}.  The matrix $\mathcal{A}\left(\chi\right)$ of \eqref{eq:eq.016.ODE.chi.p.Hamiltonian}  gives the linear, angle dependent and inertia dependent relationship between $\vec{p}$ and $\vec{\omega}$, as  
\begin{equation}
\vec{\omega}=\mathcal{A}\left(\chi\right)\cdot\vec{p}\,.
\label{reduction}
\end{equation}
Alternatively,   in terms of the time derivatives of the Euler angles, $\vec{\omega}$ can be expressed as  $\vec{\omega}=D\left(\chi\right)\cdot\vec{\dot{\chi}}$  (see Figure \ref{figure13} and \cite{Morrison98}) where 
\begin{eqnarray}
D\left(\chi\right)&=&\left(\begin{array}{ccc}
\cos\chi_{3} & \sin\chi_{1}\sin\chi_{3} & 0\\
-\sin\chi_{3} & \sin\chi_{1}\cos\chi_{3} & 0\\
0 & \cos\chi_{1} & 1
\end{array}\right),
\nonumber\\
\mathcal{A}\left(\chi\right)&=&\sigma^{-1}\cdot\left(D^{-1}\left(\chi\right)\right)^{\mathrm{T}}.
\end{eqnarray}
Using the above formulae, Equations  (\ref{eq:eq.016.ODE.chi.p.Hamiltonian}) do indeed reduce  to the equations of motion for $\vec{\omega}$ obtained by inserting $\vec{\omega}$ in (\ref{eq:dotA.Hamiltonian.L}).  Thus, there is a correspondence between the solutions of the two formulations.    The beauty of Euler's reduction is that one need not solve all six equations at once, but  one can solve first for the three components of $\vec{\omega}$,  then afterwards solve for the time dependence of the three angles $\chi$.  Because of  \eqref{omegaL}, one can use $\vec{\omega}$ or $\vec{L}$ interchangebly -- in this section we will use $\vec{\omega}$.

Consider  the particular equilibrium  angular velocity
\begin{equation}
\vec{\omega}^*=\Omega\hat{e}_{2}\,,
\label{eq:eq017.omegastar.Omega.e2}
\end{equation}
corresponding to the rigid body rotating around the $\hat{e}_2$ axis.  If the momentum $I_2$ is either the largest or the smallest eigenvalue of $\sigma$, then $\vec{\omega}^*$ in (\ref{eq:eq017.omegastar.Omega.e2}) is an asymptotically stable point in the phase space of the angular velocities. Placing (\ref{eq:eq017.omegastar.Omega.e2}) into (\ref{eq:eq.016.ODE.chi.p.Hamiltonian}), and considering $\vec{\omega}=A\left(\chi\right)\cdot\vec{p}$, we find that the corresponding velocities of the Euler angles, 
\begin{equation}
\vec{\omega}=\vec{\omega}^*\ \Rightarrow\ \begin{cases}
 & \dot{\chi}_{1}=-\Omega\, \sin\chi_{3},\\
\\
 & \dot{\chi}_{2}=\Omega\, {\displaystyle \frac{\cos\chi_{3}}{\sin\chi_{1}}},\\
\\
 & \dot{\chi}_{3}=-\Omega\, {\displaystyle \frac{\cos\chi_{1}\cos\chi_{3}}{\sin\chi_{1}}}.
\end{cases}\label{eq:eq018}
\end{equation}
A possible, particularly understandable, solution of (\ref{eq:eq018}), is\begin{equation}
\begin{cases}
 & \chi_{1}\left(t\right)=\frac{\pi}{2},\\
 & \chi_{2}\left(t\right)=\Omega t+\chi_{2}\left(0\right),\\
 & \chi_{3}\left(t\right)=0\quad \forall\  t.
\end{cases}\label{eq:eq019}
\end{equation}
The solution (\ref{eq:eq019}) represents \emph{a uniform rotation around the node line}, that will coincide with the new ``$X$-axis'' defined by  $\chi_3 = 0$ at every time. The canonical momentum  corresponding to $\vec{\omega}^*$ is the vector  $\vec{p}^*=\mathcal{A}^{-1}\left(\chi^*\right)\cdot\vec{\omega}^*$, with $\chi^*\left(t\right)=\left(\frac{\pi}{2},\Omega t+\chi_{2}\left(0\right),0\right)$, that reads
\begin{equation}
\vec{p}^*=\left(\begin{array}{c}
0\\
I_{2}\Omega\\
0
\end{array}\right).
\label{eq:eq020.p.star}
\end{equation}
This $\vec{p}^*$ and the aforementioned $\chi^*$ form a solution of the ODEs (\ref{eq:eq.016.ODE.chi.p.Hamiltonian}) in the space of canonical variables $\left(\chi,\vec{p}\right)$, as they correspond to the $\vec{\omega}^*$ solving the Hamilton equations in the $\vec{\omega}$-space.

What happens to this trajectory as it is approached under the metriplectic dynamics?  To understand this, one would need to know the pre-image of 
the terms that map into those of the symmetric bracket under the reduction map.   To understand this recall from \eqref{reduction} that 
\begin{equation}
\label{MPR}
\dot{\vec{\omega}}= \frac{\partial \vec{\omega}}{\partial \chi}\cdot \dot{\chi} + \frac{\partial \vec{\omega}}{\partial \vec{p}} \cdot \dot{\vec{p}}\,.
\end{equation}
When the canonical Hamiltonian equations of \eqref{eq:eq.016.ODE.chi.p.Hamiltonian} for $\dot{\chi}$ and $\dot{\vec{p}}$ are substituted into the above, the righthand side collapses down to a function of  $\vec{\omega}$ alone.  We seek dissipative  terms that can be added to the equations of  \eqref{eq:eq.016.ODE.chi.p.Hamiltonian} such that when we again substitute  $\dot\chi$ and $\dot{\vec{p}}$ the right hand side of \eqref{MPR} again becomes   a function of  $\vec{\omega}$ alone and adds the metriplectic dissipation terms produced by the symmetric bracket of the metriplectic dynamics of Section \ref{sec:servomotor}.  Achieveing this would be a case of  metriplectic reduction. 

Since dissipation often appears only  in the $\dot{\vec{p}}$ equation, for simplicity  we do this here, call it $\vec{\Delta}_p\left(\chi,\vec{p}\right)$,  and explore the consequences.  Comparison with the metriplectic dissipation of \eqref{MPdotL} yields the following:
 \begin{equation}
\vec{\Delta}_{p}=2\zeta C' \mathcal{A}^{-1}\cdot\Gamma\cdot\sigma^{2}\cdot \mathcal{A}\cdot\vec{p},\label{eq021.Delta.p}
\end{equation}
and the unreduced equations have the form
\begin{equation}
\begin{cases}
 \quad \vec{\dot{\chi}}&=D^{-1}\cdot \mathcal{A}\cdot\vec{p},\\
 \quad  \dot{\vec{p}}&=-\vec{p}^{\mathrm{T}}\cdot \mathcal{A}^{\mathrm{T}}\cdot\sigma\cdot{\displaystyle \frac{\partial \mathcal{A}}{\partial\chi}}\cdot\vec{p}\\
 &\quad +2\zeta C' \mathcal{A}^{-1}\cdot\Gamma\cdot\sigma^{2}\cdot \mathcal{A}\cdot\vec{p}.
\end{cases}
 \label{eq022.chidot.pdot.MSM.Deltap}
\end{equation}
Thus we have obtained an example of metriplectic reduction.

It is possible to see that the curve (\ref{eq:eq019}) and (\ref{eq:eq020.p.star}) in the full canonical phase space of the rigid body solves the total system of (\ref{eq022.chidot.pdot.MSM.Deltap}), because the metriplectic term $\vec{\Delta}_{p}$ can be shown to vanish, as it must, when evaluated on $\left(\chi^*\left(t\right),\vec{p}^*\right)$.  This is due to the fact that the matrix
\begin{equation}
\Gamma\left(\vec{\omega}^{*}\right)=\frac{k\Omega^{2}}{\zeta}\left(\begin{array}{ccc}
1 & 0 & 0\\
0 & 0 & 0\\
0 & 0 & 1
\end{array}\right)
\end{equation}
has the vector
\begin{equation}
\sigma^{2}\cdot \mathcal{A}\left(\chi^{*}\right)\cdot\vec{p}^{*}=\left(\begin{array}{c}
0\\
I_{2}^{2}\, \Omega\\
0
\end{array}\right)
\end{equation}
in its kernel. Hence, one may state that, \emph{once on this orbit, the rigid body will remain there even in the presence of metriplectic MSM torque. } 

 As mentioned before, if  $I_2$ is either the maximum or the minimum   eigenvalue of $\sigma$, the point (\ref{eq:eq017.omegastar.Omega.e2}) represents a stable equilibrium point for the Hamiltonian free rigid body, and is an asymptotically stable point for its metriplectic counterpart. If the canonical variables $\left(\chi,\vec{p}\right)$ are adopted, the corresponding $\left(\chi^*(t),\vec{p}^*\right)$ is periodic orbit  of the  system that lies in an attracting cylinder. In order to check this, we expand (\ref{eq022.chidot.pdot.MSM.Deltap}) in $\left(\delta\chi,\delta\vec{p}\right)$ around $\left(\chi^*(t),\vec{p}^*\right)$ giving rise to  the perturbations $\left(\delta\chi_2,\delta p_2\right)$ decoupling from the others, 
\begin{equation}
\delta\dot{\chi}_{2}=\frac{\delta p_{2}}{I_{2}}\quad\mathrm{and}\quad  \delta\dot{p}_{2}=0\,.
\end{equation}
Thus, 
\begin{eqnarray}
\chi_{2}\left(0\right)&\mapsto&\chi_{2}\left(0\right)+\delta\chi_{2}\left(0\right)
\nonumber\\
 \Omega &\mapsto&\Omega+\frac{\delta p_{2}\left(0\right)}{I_{2}}
\end{eqnarray}
gives rise to a neighboring cycle that is  swept starting from an adjacent point and with an adjacent  velocity.  As one moves up the cylinder (increasing $p_2$) the frequency of rotation around the cylinder increases.   Because this decoupled eigenvalue problem has zero eigenvalues,  the solution is not attracting within the cylinder but has a neighboring rotational state.

The other variables evolve according to the following  system of linear equtions: 
\begin{equation}
\begin{cases}
 & \delta\dot{\chi}_{1}=-\frac{\left(I_{1}-I_{2}\right)\Omega}{I_{1}}\delta\chi_{3}+\frac{\delta p_{1}}{I_{1}},\\
 & \delta\dot{\chi}_{3}=\Omega\delta\chi_{1}+\frac{\delta p_{3}}{I_{3}},\\
 & \delta\dot{p}_{1}=I_{2}\Omega^{2}\delta\chi_{1}+2\beta I_{2}\left(I_{1}^{2}-I_{2}^{2}\right)\Omega^{3}\delta\chi_{3}\\
 &  \quad\quad   +2\beta\left(I_{1}^{2}-I_{2}^{2}\right)\Omega^{2}\delta p_{1}+\Omega\delta p_{3},\\
 & \delta\dot{p}_{3}=\frac{I_{2}\left(I_{1}-I_{2}\right)\Omega^{2}}{I_{1}}\delta\chi_{3}+\frac{\left(I_{1}-I_{2}\right)\Omega}{I_{1}}\delta p_{1}\\
 &  \quad\quad \ +2\beta\left(I_{3}^{2}-I_{2}^{2}\right)\Omega^{2}\delta p_{3},
\end{cases} \label{eq023}
\end{equation}
where the assumption $\beta=kC'\left(\ L^2\left( \chi^* , \vec{p}^* \right) \right)$ has been made. The study of the system (\ref{eq023}) gives the expected results, viz.\  the perturbations of the limit cycle (\ref{eq:eq019}) and (\ref{eq:eq020.p.star}) tend to zero with time as $I_1 - I_2$ and $I_1 - I_3$ have the same sign, i.e.\ as the body nearly rotates around a principal axis with either maximum or minimum moment of inertia. This properly represents how the point-like equilibrium (\ref{eq:rigidbody.CMS.equilibrium}) in the reduced phase space turns into an attracting cylinder in the complete canonical phase space.

\section{Conclusions \label{sec:conclusion}}

In this paper we have briefly reviewed  metriplectic systems, the extension of Hamiltonian systems that includes dissipation that pushes the system to  asymptotically stable equilibrium points while maximizing entropy at fixed energy.   The example of a rotating rigid body was described and the control theory implication of the  metriplectic servo-motor,  where metriplectic dissipation may drag the angular velocity to align with one of its principal axes of inertia, without the dissipation of mechanical energy, was discussed.  Several numerical examples depicting this relaxation were shown.  A formulation of the rigid body metriplectic system in terms of canonical variables, the Euler angles and their conjugate momenta,  was also given and the idea of metriplectic reduction was introduced, whereby  canonical systems with appropriate dissipation are reduced to metriplectic systems.  Given the wide interest in various kinds of attractors across many fields of science, it is interesting that the unreduced rigid body system has an attracting cylinder of periodic orbits. 

There are many avenues for future work.  Given that many of the partial differential  equations that describe plasmas and fluids possess metriplectic formulations, calculations analogous to the ones done here can be done in the infinite-dimensional context for these systems.  In particular,  metriplectic reduction can be explored for these systems. Our example of metriplectic reduction was only a special case, even for our finite system,  and the general investigation of reducible systems awaits. Since the early work of \cite{Souriau} reduction has been placed in a Lie group setting.  It is quite clear that there is such a symmetry group theory interpretation of metriplectic reduction and this will the subject of future research.  


\section*{Acknowledgements}
PJM received support from the US Dept.of Energy Contract DE-FG05-80ET-53088 and from a Forschungspreis from the Alexander von Humboldt Foundation.   He would like to warmly acknowledge the hospitality of the Numerical Plasma Physics Division of the Max Planck IPP, Garching, Germany.

\bibliographystyle{physcon}

\begin{thebibliography}{}

\bibitem[Morrison, 1986] {Morrison86} Morrison, P. J (1986) {\em A Paradigm for Joined Hamiltonian and Dissipative Systems}. Physica D 18, 410.

\bibitem[Morrison, 1984a]{Morrison84a} Morrison, P. J. (1984a) {\em Bracket Formulation for Irreversible Classical Fields}. Phys. Lett. A 100, 423.

\bibitem[Morrison, 1984b]{Morrison84b} Morrison, P. J. (1984b) {\em  Some Observations Regarding Brackets and Dissipation}.  Center for Pure and Applied Mathematics Report PAM--228, University of California, Berkeley (1984).

\bibitem[Materassi and Tassi, 2012]{MateTassi12} Materassi, M., Tassi, E. (2012) {\em Metriplectic Framework for Dissipative Magneto-Hydrodynamics}. Physica D, 241, 6.

\bibitem[Morrison, 1998] {Morrison98} Morrison, P. J (1998) {\em Hamiltonian description of the ideal fluid}. Rev. Mod. Phys., 70, 467--521. 
\bibitem[Materassi, 2012] {Materassi12} Materassi M. (2012), {\em Entropy as a Metric Generator of Dissipation in Complete Metriplectic Systems}. Entropy, 18, 304; doi:10.3390/e18080304.

\bibitem[Bloch, Morrison, and Ratiu, 2013]{BMR13}  Bloch, A. M., Morrison, P. J.,   Ratiu, T. S.  (2013) {\em Gradient Flows in the Normal and Kaehler Metrics and Triple Bracket Generated Metriplectic Systems},  in Recent Trends in Dynamical Systems, eds.\  A. Johann et al., Springer Proc. in Mathematics \& Statistics 35,  371--415; doi:10.1007/978-3-0348-041-6$\_$15. 

\bibitem[Souriau, 1970]{Souriau} Souriau, J.-M. (1970), Structure des syst\`{e}mes dynamiques (Dunod, Paris).


\end{thebibliography}

\end{document}